# Encoding *M* classical bits in the arrival time of dense-coded photons

Salem F. Hegazy[a,c], Ahmed E. Morra[b,c], and Salah S. A. Obayya[c*]

[a]National Institute of Laser Enhanced Sciences, Cairo University, 12613 Giza, Egypt
[b]Faculty of Electronic Engineering (FEE), Menoufia University, 32952 Menouf, Egypt
[c]Centre for Photonics and Smart Materials, Zewail City of Science and Technology, 12588 Giza, Egypt

## ABSTRACT

We present a scheme to encode *M* extra classical bits to a dense-coded pair of photons. By tuning the delay of an entangled pair of photons to one of $2^M$ time-bins and then applying one of the quantum dense coding protocols, a receiver equipped with a synchronized clock of reference is able to decode *M* bits (via classical time-bin encoding) + 2 bits (via quantum dense coding). This protocol, yet simple, does not dispense several special features of the used programmable delay apparatus to maintain the coherence of the two-photon state. While this type of time-domain encoding may be thought to be ideally of boundless photonic capacity (by increasing the number of available time-bins), errors due to the environmental noise and the imperfect devices and channel evolve with the number of time-bins.

## 1. INTRODUCTION

Superdense coding shows one of the mysterious features of quantum information processing[1]. Superdense coding is one among several quantum schemes where classical information capacity is investigated for noisy quantum channels. The classical encoding is accomplished by the transmitter "Alice" choice of a quantum state among a predefined set of states to be sent to the receiver "Bob". For one-way transfer, the maximum mutual information, or so called the information capacity, is bounded by the associated Hilbert-space dimension[2]. Superdense coding uses a preshared quantum resource so that Alice can signal Bob at a mutual information rate beyond (up to twice when sharing maximally entangled state[3]) the capacity bound given by its Hilbert-space dimension, yet still satisfying the upper bound of quantum-information causality[4].

For example, the photonic version of the original discrete-variable superdense coding scheme[1,5] uses a polarization qubit to transfer two classical bits from Alice to Bob who previously shared polarization-entangled photons. While Bob is theoretically able to determine the 2 bits using Bell state analysis (BSA), experimentally, linear optics can only discriminate three states out of four, limiting the channel capacity to $\log_2 3 \approx 1.585$ bits[5] [mattle96]. By virtue of hyperentanglement, linear optics is not only able to perform fully deterministic BSA retrieving, in principle, the 2 bits/qubit[6-8] (Kwiat 1998, chuck2006, Barbierie2007), but also capable of distinguishing 7 classes of 16 hyper-Bell spin-orbit states rising the maximum channel capacity to $\log_2 7 \approx 2.81$ bits[9,10] (Wei 2007; Barreiro 2008). In other words, the multiple degrees of freedom by exploiting hyperentanglement offer larger coding space or denser coding.

On the other hand, the modulation in different degrees of freedom, or so called hybrid modulation, is well known in classical optical communications. Its advantages span from reducing the required signaling photons per bit, or equivalently increasing the power efficiency, to enhancing the spectral efficiency of the transmission. Recently, the key trend of hybrid modulation is to use a spectrally efficient modulation scheme, such as binary phase shift keying (BPSK), polarization-division-multiplexed quadrature phase-shift keying (PDM-QPSK), and differential phase shift Keying (DPSK), along with an energy-efficient modulation scheme, such as M-ary pulse-position modulation (M-ary-PPM) and multipulse pulse position modulation (MPPM), so as to integrate the advantages of both schemes[11]. M-ary-PPM and so hybrid modulation techniques are capable of achieving Shannon's limit by simply increasing the number of time slots ($M$) when used with ideal photon-counting receivers[12].

---

[*]sobayya@zewailcity.edu.eg

The excessive capacity realized by superdense coding can be interpreted in a likely manner. While the nonlocal correlations offered by entanglement in discrete variables (e.g., polarization and spin) or continuous variables (e.g., position-momentum and energy-time) obey the no-signaling principle (That is a measurements outcome by one side do not reduce his uncertainty regarding the actions made by the other side), it appears as a "capacity-doubling" resource accessible by an appropriate superdense coding scheme[1,3,13]. However, we believe that a hybrid scheme that integrates a quantum superdense coding and a classical energy-efficient modulation (like PPM) can offer much higher information capacity.

In this paper, we propose a scheme to encode $M$ classical bits by tuning the arrival time of dense-coded Einstein-Podolsky-Rosen (EPR) photons to one of $2^M$ distinguishable time-bins. In contrast to the known time-bin qubit described by the superposition of two time modes $|early\rangle \equiv |0\rangle$ and $|late\rangle \equiv |1\rangle$, the $M$-bits time-bin symbol, that we mention here, takes its value based on the deterministic detection-time of the EPR pair among $2^M$ prior choices. Classical time-bin bit in this sense is an additional degree of freedom that the information can span beside other quantum degrees of freedom like polarization, phase, spatial mode and momentum.

By means of an ultra-fast discrete-delay line, Alice, who initially owns a pair of degenerate polarization-entangled photons, is capable of introducing a time delay $iT_b$ ($i = 1,2,\ldots 2^M$) to the pair before being sent to Bob (where the time-bin interval $T_b$ is sufficiently longer than $t_c$ the coherence time of the entangled photons upon spectrally-selective detection at Bob side). In such way, the time-encoded $M$ classical bits can be set well beyond the current photon capacity record achievable by making use of the orbital angular momentum qubit (2.81 bits per photon). However, higher bit rate per photon in ideal conditions is associated with higher quantum bit error rate (QBER) in actual case. Bob's detectors listening to Alice's pair for more vacant (non-signal) time-bins are higher subjected to noise accidental counts either due to environmental or dark photon counts. It is therefore a tradeoff and the photon capacity is upper-bounded due to the QBER.

## 2. CLASSICAL TIME-BIN CODING OF EPR PHOTONS

Consider that Alice, who periodically prepares (or owns) a pair of photons of the polarization-entangled state

$$|\Psi^+\rangle = \frac{1}{\sqrt{2}}\{|H\rangle_1|V\rangle_2 + |V\rangle_1|H\rangle_2\}, \quad (1)$$

can coherently control the time of their transport to Bob by means of a programmable discrete delay line to be one of $2^M$ time-bins, as depicted in Fig. 1. Here, we mean by the coherent delay that it does not disturb the entangled state in (1)

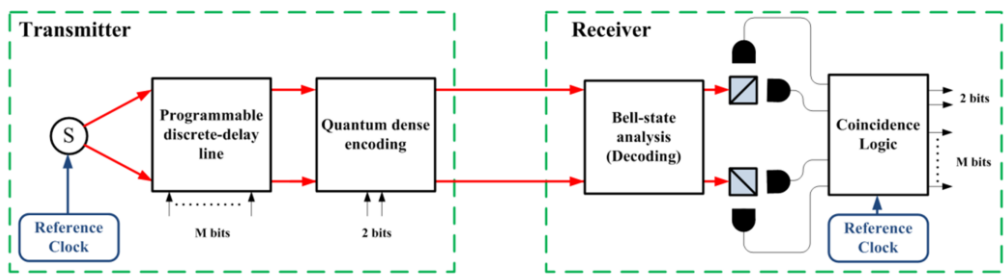

Figure 1. Schematic for the proposed quantum encoding system

and that the output state reads

$$|\Psi^+, t_i\rangle = \frac{1}{\sqrt{2}}\{|H, t_i\rangle_1 |V, t_i\rangle_2 + |V, t_i\rangle_1 |H, t_i\rangle_2\}, \qquad i = \{1, 2, \dots, 2^M\} \qquad (2)$$

Alice then applies a scheme of quantum dense coding by acting on one of the two photons, choosing one of the four maximally entangled states, or so called Bell states $\{|\Psi^\pm\rangle, |\Phi^\pm\rangle\}$. Here, Alice might exploit the entanglement in an additional degree of freedom to enable the full discrimination of Bell states at Bob side using linear optics. Alice's biphoton therefore spans a space of $2^{(M+2)}$ dimensions that enables the coding of $(M + 2)$ bits per biphoton. The photon pair then is transported down the optical channel between Alice and Bob.

Bob runs an appropriate BSA to deterministically decide which of the four Bell states was sent. Bob's receiving station is equipped by a reference clock signal which is precisely synchronized to that at Alice's. Therefore after the two-photon measurements, Bob does not only register which couple of the four detectors simultaneously ticks (gives the 2 bits due to superdense coding), but also when this coincidence takes place with reference to the local clock of reference (gives the $M$ bits of the time-bin encoding). Bob can therefore determine the $(M + 2)$ bits encoded by Alice.

### 3. COHERENCE-MAINTAINING PROGRAMMABLE DELAY LINE

The coherent delay of the biphoton is not a straightforward task. In the following, we mention about the main aspects required in the delay line of the EPR photons.

It is, firstly, required for such a discrete-delay line to be polarization independent, i.e., the delay is invariant with the polarization mode. This is essential to maintain the temporal indistinguishability of the two-photon wavepacket.

Secondly, it is required that the delay line does not build up any correlations between the polarization and any other degree of freedom (e.g., spatial mode, frequency), thereby the degree of entanglement is conserved. Actually, the delay line should be designed to wash any correlations away, if applicable.

Thirdly, the introduced delay should be precisely the same for the pair of photons. Specifically, the differential delay (if exists) should be kept sufficiently less than the coincidence time window of the receiver logic (which, in turn, is appropriately less than the period between two sequential time-bins).

Fourthly, the delay-dependent phase received along the traverse down the delay line should not disturb the relative phase of the maximally entangled state in (1). Relative phase distortions can flip the state in (1) to another orthogonal maximally entangled state. Since we plan for a hybrid dense coding technique where time-bin encoding coexists with traditional dense coding, distortions of the relative phase will directly induce errors in the traditional dense coding.

Fifthly, the two-photon should exit the delay line in two distinguishable spatial modes (again, without correlations with polarization) to enable the operation of the subsequent dense coding at Alice's station and the BSA at Bob's station.

Historically, there have been two main structures for programmable discrete optical delay lines based on fiber coils, namely, the square-root cascaded delay line[14] (SRODEL) and the binary fiber-optic delay line[15] (BIFODEL). Both approaches rely on switching the light among a number of fiber coils of different lengths. Relative to the SRODEL, the BIFODEL is more compact with utilization ratio 1:2 (i.e., half of the fiber segments are used for each case), as depicted in Fig. 2.

One should notice that using a separate delay line for each photon will make the two-photon state heavily subjected to distortions due to the relative dynamical changes in their delay and phase. It is then suitable to use a fiber-optic delay line that allows a single spatial mode of propagation for the two-photon. Besides the known advantage of avoiding the spatial mode dispersion (naturally occurs in multimode fiber), single-mode propagation restricts the phase accumulated along the delay line to one stable relation (given that the fiber is not elongated). Further, single-mode propagation eliminates the correlations with the spatial mode, thereby limiting the "which-path" information.

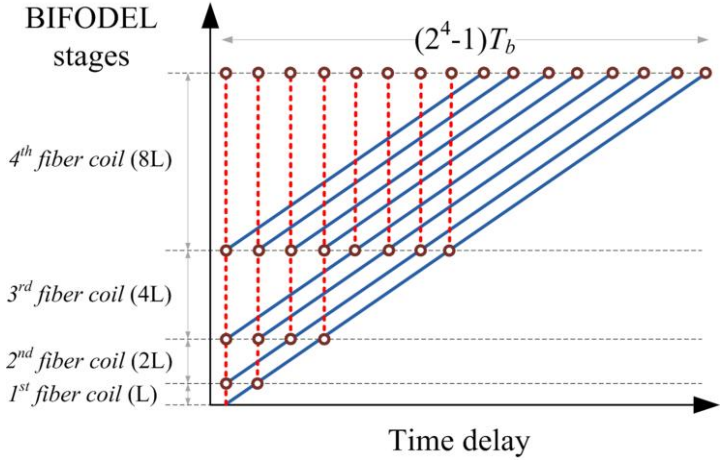

Figure 2. Four-stage BIFODEL with up to 15 delay steps (time-bins). The blue-solid and the red-dashed lines denote light traversing longer or shorter fiber routes, respectively. The circles mark the available delay steps after each stage.

However, single-mode fiber is not an ideal environment. Slight birefringence always exists due to either geometrical asymmetries or the residual stress in core or cladding. Therefore, the state of polarization (SOP) along the fiber unitarily slides over the Poincare sphere (representing the SOP space). Polarization controller per each fiber coil can passively recover the initial SOP provided that there is no mechanical or thermal fluctuations (active birefringence compensation can be carried out by retracing the beam back through the fiber coil[16].) Moreover, due to the weak birefringence, the two orthogonal polarization modes easily couple to each other resulting in what so called polarization mode dispersion.

Another BIFODEL approach can be by using polarization-maintaining (PM) fiber where its heavy birefringence prohibits coupling between the two orthogonal polarization modes[11] (therefore, no need for the polarization controller). PM fiber has a significant differential group delay that takes place between the two modes of propagation allowing fast and slow routes per fiber coil. Therefore, a single PM fiber coil can play the role of long and the role of short transport media.

This BIFODEL is then a series of optic-axes-aligned PM fibers which allow single-mode propagation. The optical switches between each two fiber coils should -on-demand- toggle the SOP between fast and slow passes or to leave it as it is. On-shelf ultrafast polarization modulators (up to 50 GHz) that allow single-mode guided operation can then be used for this purpose.

## 4. QUANTUM BIT ERROR RATE CALCULATION

Let us assume that the dark and environmental single-photon counts detected per time-bin follows a Poisson distribution at the receiver detectors. The symbol represents the $M$-bits word encoded in the time-bin occupied by the two-photon among $\Upsilon = 2^M$ available time-bins. The symbol error rate is then calculated following an approach similar to that adopted by Hamkins et al.[17] and Morra et al.[18] in the case of multi-pulse Pulse position modulation (MPPM) technique.

Let $K = \{K_1, K_2, \ldots, K_\Upsilon\}$ denote the coincidence counts detected along the $\Upsilon$ time-bins, and the encoding time-bin is in a position $a$ within the symbol frame. Therefore, we have one time-bin encoded by the EPR pair, $l$ vacant time-bins with accidental coincidence counts due to noise, and $(\Upsilon - l - 1)$ vacant time-bins without counts. Accordingly, There are two possible cases that may lead to a symbol error

$$\begin{cases} \text{Case A:} & (K_a \geq 1) \text{ and } (l \geq 1) \\ \text{Case B:} & (K_a = 0) \text{ and } (l \geq 1) \end{cases}$$

For case A (the encoding time-bin is nonvacant, yet other $l$ time-bins is nonvacant also), the probability of symbol error reads

$$P(e)_A = \sum_{l=1}^{Y-1} \Pr[\textit{Error Decision}]$$
$$\times \Pr[l \text{ of } Y-1 \text{ vacant time bins have a joint detection event, all others have zero counts}]$$
$$\times \Pr[\textit{occupied time bin has a joint detection event (not necessarily due to the EPR pair)}] \quad (3)$$

where Pr[*Error Decision*] arises because, for a given $l$, Bob should choose from $(l+1)$ destinct fair decisions, so that

$$\Pr[\textit{Error Decision}] = \frac{l}{l+1}$$

For case B (the encoding time-bin is vacant and $l$ time-bins is nonvacant), the probability of symbol error is

$$P(e)_B = \sum_{l=1}^{Y-1} \Pr[l \text{ of } (Y-1) \text{ vacant time bins have a joint detection event, all others have zero counts}]$$
$$\times \Pr[\textit{occupied time bin has zero counts}] \quad (4)$$

Therefore, the total symbol error rate is given as

$$P_{eSym} = P(e)_A + P(e)_B \quad (5)$$

We can express the quantum bit-error rate (QBER) in this case in terms of the symbol-error rate[19] as

$$\text{QBER} = \frac{Y}{2(Y-1)} P_{eSym} \quad (6)$$

One can conclude that as the number of time-bins $Y$ or the rate of noisy photons $R$ increase [which can be thought as less signal-to-noise (SNR) ratio, in analogy to classical communications] , the QBER should grow up as well. Figure 3 shows the QBER versus ϒ and $R$. It can be observed that by increasing either ϒ or $R$, the QBER of the proposed system increases as expected.

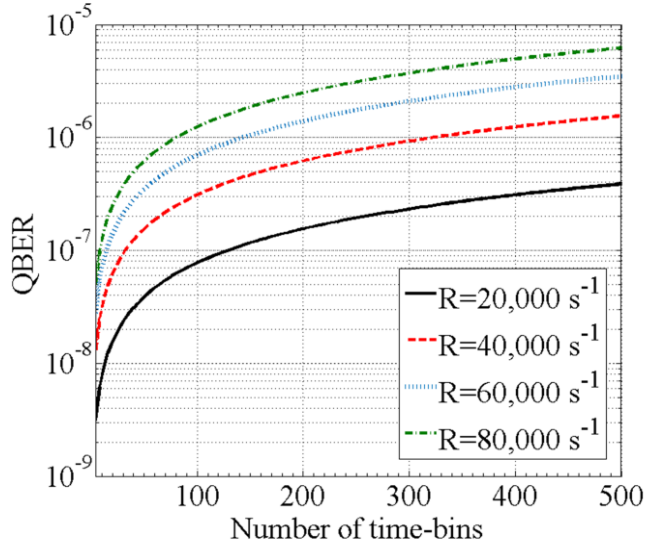

Figure 3. QBER versus number of time-bins per symbol $Y$ under different values of background single-photons rate $R$. Here, we assume that the quantum efficiency of each detector is 50%, the channel and receiver-optics loss is 50%, and the coincidence window is 1 ns.

## ACKNOWLEDGEMENTS

This work was supported by the Information Technology Industry Development Agency (ITIDA-ITAC), Ministry of Communication, Egypt.